\documentstyle[12pt]{article}
\begin{document}
\begin{center}
{\Large\bf Path-ordered Phase Factors in Scalar}
\vspace{0.2cm}

{\Large\bf Quantum Electrodynamics}
\vspace{1.0cm}\\
B. M. Pimentel\footnote{Partially supported by CNPq} and J. L.
Tomazelli\footnote{Supported by CAPES}
\vspace{0.2cm}

Instituto de F\'{\i}sica Te\'{o}rica \\
Universidade Estadual Paulista \\
Rua Pamplona, 145 \\
01405--900 - S\~{a}o Paulo, SP - Brazil

\vspace{1.0cm}

{\large\bf Abstract}
\end{center}
\vspace{0.3cm}

Starting from linear equations for the complex scalar field, the two- and
three-point Green's functions are obtained in the infrared approximation. We
show that the infrared singularity factorizes in the vertex function as in
spinorial QED, reproducing in a straightforward way the result of lenghty
perturbative calculations.

\vspace{0.4cm}

electromagnetic interactions.

\vspace{1.0cm}

{\large\bf 1.\,Introduction}
\vspace{0.5cm}

The major difficulty in dealing with long-range potentials in the
relativistic quantum theory is the factorization of infrared divergences out
of scattering amplitudes, in order to get finite cross-sections for the
corresponding processes. Who first provided a deeper insight to this problem
in spinorial quantum electrodynamics were Bloch and Nordsieck${}^{[1]}$, over
half a century ago. Later, it was shown${}^{[2], [3]}$ that if we cast
bremsstrahlung contributions to all orders in perturbation theory, we get
finite cross-sections. Pursuing the Bloch and Nordsieck lines,
Chung${}^{[4]}$, Kulish and Faddeev${}^{[5]}$, Zwanziger${}^{[6]}$, and
others work in the construction of a new space of asymptotic states,
containing an infinite number of coherent photons, in order to extract finite
elements from the S-matrix. Nevertheless, despite these great efforts, one
feels that few progress have been made along the last decades at a more
operational level, and even more puzzling is the case of non-Abelian gauge
theories, such as quantum chromodynamics, and quantum gravity.
In the early 80's, applying the concept of path-ordered phase factors,
exhaustive explored in QCD, Kubo${}^{[7]}$ reproduced the Kulish-Faddeev and
Grammer-Yennie results for spinorial QED in the infrared asymptotic region.
The power of his method lies on the simplicity of calculations. Furthermore,
it can be extended to the determination of generic n-point Green's functions.
It would be desirable that Kubo's method also applies to scalar QED in order
to obtain information about its behavior at the infrared, since few emphasis
has been given to this matter and the analysis is much less obvious in this
case. As we shall see, it is sufficient to write the interaction Hamiltonian
of scalar QED in a suitable form.

Keeping this in mind, we will adopt an alternative approach to scalar QED,
which is also useful for the investigation of electromagnetic properties of
particles of either zero or unity spin. It is specially adequate to the study
of the renormalizability of scalar QED as an effective theory${}^{[8]}$,
equivalent to the original, whose structure is similar to that of spinorial
QED. This formalism is briefly described in section 2. In section 3 we give
the asymptotic form for the current of charged mesons. In the sequence,
sections 4 and 5, we evaluate the dressed meson propagator and the vertex
function in the infrared region and confront the results with those obtained
through different methods.

\newpage

{\large\bf 2.\, The Duffin-Kemmer equation for the scalar field}
\vspace{0.5cm}

For our purposes it is convenient to start not with the usual Klein-Gordon
field equation but with the first order equation
\begin{equation}
(i\beta_{\mu} \partial^{\mu}-m) \psi(x)=0 \,\,,
\end{equation}
known as the Duffin-Kemmer equation${}^{[9]}$, which has the same appearance
as the Dirac equation, where $\beta_{\mu}$ are matrices satisfying
the algebra
\begin{equation}
\beta_{\mu} \beta_{\nu} \beta_{\lambda}+\beta_{\lambda} \beta_{\nu} \beta_{\mu}
=g_{\mu \nu} \beta_{\lambda}+g_{\lambda \nu} \beta_{\mu} \,\,\,,
\end{equation}
with metric tensor $g_{\mu \nu}=diag(1,-1,-1,-1)$. In the linear equation (1)
the field $\psi$ has five components, the scalar field and its four
derivatives with respect to the coordinates and the time which in turn
transform like the components of a vector.

The adjoint $\overline{\psi}(x)$ of $\psi(x)$ is defined by
\begin{equation}
\overline{\psi}(x)=\psi^{\dagger}(x)(2{\beta_{0}}^2-I)\,\,\,,
\end{equation}
where $I$ is the five-rowed identity matrix. Using that the matrix
$\beta_{0}$ is Hermitian and the matrices $\beta_{k}$ are anti-Hermitian for
$k=1,2,3$, it can be shown that $\overline{\psi}(x)$ satisfies
\begin{equation}
i\partial^{\mu} \overline{\psi}(x) \beta_{\mu}+m\overline{\psi}(x)=0 \,\,\,.
\end{equation}
Equations (1) and (4) are obtained from the Lagrangian density
\begin{equation}
{\cal L}_{0}^m=\frac{i}{2}(\overline{\psi} \beta_{\mu} \partial^{\mu}\psi-
\partial^{\mu}\overline{\psi} \beta_{\mu} \psi )-m\overline{\psi}\psi \,\,\,,
\end{equation}
and the corresponding charge-current vector for the scalar field can be
written as
\begin{equation}
j_{\mu}=e\overline{\psi} \beta_{\mu} \psi \,\,\,.
\end{equation}

The sistem of the meson field and the electromagnetic field in interaction is
classically described by the Lagrangian density
\begin{equation}
{\cal L}={\cal L}_{0}^m+{\cal L}_{0}^{em}+{\cal L}_{int} \,\,\,,
\end{equation}
where ${\cal L}_{0}^{em}$ represents the Lagrangian density for the free
electromagnetic field and
\begin{equation}
 {\cal L}_{int}=j_{\mu}A^{\mu}
\end{equation}
the interaction Lagrangian density. Thus, the Heisenberg equations of motion
for interacting fields are given by
\begin{eqnarray}
[i\beta^{\mu} (\partial_{\mu}-ieA_{\mu}(x))-m] \psi(x)=0 \,\,,\\
\overline{\psi}(x)[i\beta^{\mu}({}^{{}^{{}^{\leftarrow}}}\!\!\!\!\partial_{\mu}
-ieA_{\mu}(x))+m]=0\,\,,\\
\Box A_{\mu}(x)=j_{\mu}(x)\,\,.
\end{eqnarray}
The operator ${}^{{}^{{}^{\leftarrow}}}\!\!\!\!\partial_{\mu}$ in equation
(10) acts to the left by definition.

In spinor quantum electrodynamics the field operators $\psi(x)$ and
$A_{\mu}(x)$ in the interaction picture have the same form as the operators
for the corres\-ponding free fields. The situation is different here, since
not all the components of $\psi(x)$ are dynamically independent. This can be
seen if we multiply Eq.(9) to the left by $I-\beta_{0}^2$ and use relation
(2), arriving at
\begin{equation}
(I-\beta_{0}^2)\psi(x)=-\frac{i}{m}\vec{\beta}.(\stackrel{\longrightarrow}
{\nabla} -ie\vec{A}(x))\beta_{0}^2\psi(x)\,\,.
\end{equation}
In view of this subsidiary condition for the field operators $\psi(x)$, the
interaction Hamiltonian density turns out to be
\begin{equation}
{\cal H}_{int}(x)=-e\overline{\psi}^{(0)}(x)\beta^{\mu}A_{\mu}^{(0)}(x)[I+
\frac{e}{m}(I-\beta_{0}^2)\beta^{\nu}A_{\nu}^{(0)}(x)]\psi^{(0)}(x) \,\,,
\end{equation}
where ${\psi}^{(0)}(x)$ and $A_{\mu}^{(0)}$ are the operators for the free
meson and electromagnetic fields, respectively.

The vacuum expectation value of the chronological product of
${\psi}^{(0)}(x)$ and $\overline{\psi}^{(0)}(y)$ is
\begin{equation}
<0|T\{{\psi}^{(0)}(x)\overline{\psi}^{(0)}(y)\}|0>
=G^{c}(x-y)+\frac{1}{m}(I-\beta_{0}^2)\delta^{(4)}(x-y) \,\,,
\end{equation}
where $G^{c}(x-y)$ is the Green's function for the free meson field equation,
namely
\begin{eqnarray}
G^{c}(x-y)=\frac{1}{(2\pi)^4}\int d^4p\,e^{-ip(x-y)}S_{F}(p) \,\,, \\
S_{F}(p)={(\slash\!\!\!p-m)}^{-1}=
\frac{\slash\!\!\!p(\slash\!\!\!p+m) +p^2-m^2}{m(p^2-m^2+i\epsilon)} \,\,,
\end{eqnarray}
with $\slash\!\!\!p \equiv \beta_{\mu}p^{\mu}$.

We note that the Hamiltonian density (13) contains a term of order $e^2$.
Therefore, the S-matrix expansion in terms of ${\cal H}(x)$ will not be an
expansion in powers of $e$. Fortunately, the second term of (14) gives a
contribution to the S-matrix which just cancel all effects caused by the term
of order $e^2$ in the interaction Hamiltonian density${}^{[8]}$. As a result
we can evaluate S-matrix elements from the effective Hamiltonian
\begin{eqnarray}
{\cal H}_{int}^{eff}(x)=-j_{\mu}^{(0)}(x)A^{\mu (0)}(x) \,\,, \\
j_{\mu}^{(0)}(x)=e\overline{\psi}^{(0)}(x)\beta_{\mu}{\psi}^{(0)}(x) \,\,,
\end{eqnarray}
with effective pairing between meson field operators which differs from (14)
by the absence of the last term.

\vspace{1.0cm}

{\large\bf 3.\, The infrared approximation}
\vspace{0.5cm}

We may express the free field operators $A_{\mu}^{(0)}$, $\psi^{(0)}$ and
$\overline{\psi}^{(0)}$ in terms of creation and annihilation operators:
\begin{eqnarray}
A_{\mu}^{(0)}(x)={\displaystyle \frac{1}{{(2\pi)}^{\frac{3}{2}}}}\int \frac
{d^3\vec{k}}{\sqrt{2k_0}}[a_{\mu}(\vec{k})e^{-ikx}+a_{\mu}^{\dagger}(\vec{k})e^
{ikx}] \,\,, \\
\psi^{(0)} (x)={\displaystyle \frac{1}{{(2\pi)}^{\frac{3}{2}}}}\int d^3\vec{p}
\sqrt{\frac{m}{2p_0}} [b(\vec{p})u(\vec{p})e^{-ipx}+d^{\dagger}
(\vec{p})v(\vec{p})e^{ipx}] \,\,, \\
\overline{\psi}^{(0)}(x)={\displaystyle \frac{1}{{(2\pi)}^{\frac{3}{2}}}}\int
d^3 \vec{p'}\sqrt{\frac{m}{2p'_0}} [d(\vec{p'})\overline{v}(\vec{p'})e^
{-ip'x}+b^{\dagger}(\vec{p'})\overline{u}(\vec{p'})e^{ip'x}] \,\,.
\end{eqnarray}
In order to obtain an expression for the interaction Hamiltonian
\begin{equation}
H_{int}=-e\int d^3\vec{x} :\overline{\psi}^{(0)}(x)\beta_{\mu}{\psi}^{(0)}(x):
A^{\mu (0)}(x)
\end{equation}
in the interaction picture, it's sufficient to substitute formulas (19)-(21)
into the above equation. The resulting expression is an integral over the
momenta $\vec{p}$, $\vec{p'}$ and $\vec{k}$ of mesons and photons, which are
related by $\vec{p'}=\vec{p}+\vec{k}$.

In the neighborhood of small momentum $k_{\mu}$ we have
\begin{equation}
p_0-p'_0\mp k_0 \sim \mp\frac{p.k}{p_0}\,\,,
\end{equation}
and
\begin{equation}
\overline{u}(\vec{p'})\beta_{\mu}u(\vec{p})=
\overline{v}(\vec{p'})\beta_{\mu}v(\vec{p}) \sim \frac{p_{\mu}}{m} \,\,.
\end{equation}
The last result follows from an identity which can be deduced in much the same
way as the Gordon identity, using the algebra for three beta matrices, and
Dirac-like orthonormality conditions. Thus, in the infrared asymptotic region
the interaction Hamiltonian turns out to be
\begin{equation}
H^{as}_{int}=-{\displaystyle \frac{1}{(2\pi)^{\frac{3}{2}}}}\int d^3 \vec{p}\,
\frac{p_{\mu}}{p_0}\rho(\vec{p}) \int \frac{d^3 \vec{k}}{\sqrt{2k_0}}[a_{\mu}
(\vec{k})e^{-i(pk/p_0)t}+a_{\mu}^{\dagger}(\vec{k})e^{i(pk/p_0)t}] \,\,,
\end{equation}
where
\begin{equation}
\rho(\vec{p})=e[b^{\dagger}(\vec{p})b(\vec{p})-d^{\dagger}(\vec{p})d(\vec{p})]
\end{equation}
is the charge-density operator, which clearly satisfies the commutation
relations
\begin{equation}
[b(\vec{q})\,,\,\rho (\vec{p})]=b(\vec{q})\delta^{(3)}
(\vec{q}-\vec{p}) \,\,,
\end{equation}
\begin{equation}
[d^{\dagger}(\vec{q})\,,\,\rho (\vec{p})]=d^{\dagger}(\vec{q})\delta^{(3)}
(\vec{q}-\vec{p}) \,\,.
\end{equation}

By requiring that the asymptotic interaction Hamiltonian density be of the
form (17), we obtain from (19) and (25)
\begin{equation}
j_{\mu}^{as}(x)=\int d^3 \vec{p} \, \frac{p_{\mu}}{p_0}\rho(\vec{p})
\delta^{(3)}(\vec{x}-\frac{\vec{p}}{p_0}t)\,\,,
\end{equation}
for the asymptotic charged-meson current operator in the interaction picture.
This corresponds to the current of a point charge with momentum $p_{\mu}$ in a
classical uniform motion with velocity $p_{\mu}/p_0$, if we consider
charged-particle states. This coincides with the Kulish-Faddeev
result${}^{[5]}$ for fermions, thus reinforcing their statement that
expression (29) is quite general and applies in the case of charged particles
with arbitrary spin.

\vspace{1.0cm}

{\large\bf 4.\, Infrared Phase Factors}
\vspace{0.5cm}

We now apply Kubo's technique${}^{[7]}$ to extract infrared singularities
from Green's functions in the effective scalar meson theory described in
section 2. So, let us first consider the simplest case of the two-point
Green's function $G^{c}(x-y)$ given by (15). In the interaction picture, this
can be written as
\begin{equation}
G^{c}(x-y)=\frac{<0|T\{\psi^{(0)}(x)\overline{\psi}^{(0)}(y){\cal S}\}|0>}
{<0|{\cal S}|0>}\,\,,
\end{equation}
where $\psi^{(0)}$ and $\overline{\psi}^{(0)}$ are defined in (20) and (21),
respectively, with
\begin{equation}
{\cal S}=T\{\exp[i\int d^4x'\, j_{\mu}^{(0)}(x')A^{\mu (0)}(x')]\} \,\,.
\end{equation}

As we are mostly interested in the infrared asymptotic behavior of the
theory, we replace $j_{\mu}^{(0)}$ in (31) by expression (29) for
$j_{\mu}^{as}$ and neglect contributions from vacuum polarization graphs,
like in the Bloch-Nordsieck model${}^{[1]}$, since there are no antiparticles
in the limit of low frequencies. As a result the two-point Green's function
can be brought into the form
\begin{equation}
G^{c}(x-y)=<0|T\{\psi^{(0)}(x)\exp[i\int_{y_0}^{x_0} d^4x'\,
j_{\mu}^{as}(x')A^{\mu (0)}(x')]\overline{\psi}^{(0)}(y)\}|0>\,\,,
\end{equation}
where we have set the vacuum expectation value of the S-matrix equal to unity.
Taking the commutation relations (27) and (28) into account, it follows from
(15) that the Fourier transform of $G^{c}(x-y)$ is equal to
\begin{equation}
G^{c}(p)=<0|T\{\exp[ie\int_{y}^{x} dx_{\mu}'
\,A^{\mu (0)}(x')]\}|0>S_{F}(p) \,\,,
\end{equation}
where $S_{F}(p)$ is the Feynman propagator for charged mesons, given by (16),
and
\begin{equation}
x_{\mu} \equiv \frac{p_{\mu}}{p_0}x_{0} \,\,\,,\,\,\,
y_{\mu} \equiv \frac{p_{\mu}}{p_0}y_{0} \,\,.
\end{equation}

Expanding the exponential and using Wick's theorem, the path-ordered phase
factor in (33) becomes

${\displaystyle <0|T\{\exp[ie\int_{y}^{x} dx_{\mu}'\,A^{\mu (0)}(x')]\}|0>}$
\begin{equation}
=\exp\{\frac{ie^2}{2} \int_{y}^{x} dx_{\mu}' \int_{y}^{x}
dx_{\nu}''D_{c}^{\mu \nu}(x'-x'')\} \,\,,
\end{equation}
where
\begin{eqnarray}
D_{c}^{\mu \nu}(x'-x'') &=& i<0|T\{A^{\mu}(x')A^{\nu}(x'')\}|0> \nonumber \\
                        &=& {\displaystyle \frac{1}{{(2\pi)}^n}}
			\int d^nk \, e^{-ik(x'-x'')}D_{F}^{\mu \nu}(k) \,\,,
\end{eqnarray}
with
\begin{equation}
D_{F}^{\mu \nu}(k)=\frac{g^{\mu \nu}}{k^2+i\epsilon}-(1-a)
\frac{k_{\mu}k_{\nu}}{{(k^2+i\epsilon)}^2} \,\,,
\end{equation}
$a$ being the gauge parameter. We leave open the dimension in the Fourier
transform (36) to advance that we are going to employ dimensional
regularization for divergent integrals to appear in the next steps of
calculation. Inserting (36) into expression (35) and carrying out the
integrations with respect to $x_{\mu}'$ and $x_{\nu}''$, we are left with

${\displaystyle <0|T\{\exp[ie\int_{y}^{x} dx_{\mu}'\,A^{\mu (0)}(x')]\}|0>}$
\begin{equation}
=\exp \{\frac{-ie^2}{2} \int\, \frac{d^nk}{{(2\pi)}^n}\frac{1}{{(k.p)}^2}
[e^{i(\frac{x_0-y_0}{p_0})k.p}+e^{-i(\frac{x_0-y_0}{p_0})k.p}-2]
\,p_{\mu}D_{F}^{\mu \nu}(k)p_{\nu} \,\,.
\end{equation}

It is more convenient to rewrite the above equation in the form
\begin{equation}
<0|T\{\exp[ie\int_{y}^{x} dx_{\mu}'\,A^{\mu (0)}(x')]\}|0>=
e^{-f(\frac{x_0-y_0}{p_0})} \,\,,
\end{equation}
where
\begin{equation}
f(\nu)=-ie^2 \int \frac{d^nk}{{(2\pi)}^n}\,p_{\alpha}D_{F}^{\alpha \beta}(k)
p_{\beta}\int_0^{\nu} d\nu '\, \int_0^{\nu '} d\nu ''\, e^{i\nu ''(k.p)} \,\,.
\end{equation}
The last expression also appears in the calculation of the Green's function
in the Bloch-Nordsieck model for spinorial QED${}^{[10],[11]}$. Here we only
transcribe the final result for $f(\nu)$, after performing the k-integration:
\begin{equation}
f(\nu)={(-i)}^n\frac{e^2}{8{\pi}^{n/2}}\frac{\Gamma(n/2-1)}{3-n}[2p^2
+(1-a)(n-3)]P(\nu) \,\,,
\end{equation}
with
\begin{equation}
P(\nu) \equiv -{(-i\epsilon)}^{3-n}\nu+\frac{1}{4-n}[{(\nu-i\epsilon)}^{4-n}-
{(-i\epsilon)}^{4-n}] \,\,.
\end{equation}
For $n=4-\eta$, we obtain in the limit $\eta \rightarrow 0$
\begin{equation}
f(\nu)=-\frac{e^2}{8{\pi}^2}(2p^2+1-a)\{-i\frac{\nu}{\epsilon}+
\log[\frac{(i\nu+\epsilon)}{\epsilon}]\} \,\,.
\end{equation}

We now investigate the infrared singularity wich occurs in the unrenormalized
vertex function ${\cal B}_{\mu}(p,p')$, by considering the three-point
Green's function
\begin{equation}
G_{\mu}(x,y,z)=\frac{1}{{(2\pi)}^8}\int \,d^4p \int \,d^4p'
e^{-ip(x-y)-ip'(y-z)}G_{\mu}(p,p')
\end{equation}
in the interaction picture. It can be shown in the same grounds as above that
\begin{equation}
G_{\mu}(p,p')=S_{F}(p)\beta_{\mu}S_{F}(p')<0|T\{\exp [ie\int_{z}^{x}
\,dx_{\mu}'A^{\mu (0)}(x')]\}|0> \,\,,
\end{equation}
where
\begin{equation}
x_{\mu} \equiv \frac{p_{\mu}}{p_0}x_{0} \,\,\,,\,\,\,
z_{\mu} \equiv \frac{p_{\mu}'}{p_0'}z_{0} \,\,,
\end{equation}
and the integration path is along the trajectory described by the charged
meson. For convenience, we take the path from z to y and then, from y to x,
and put $y=0$. Thus, the path-ordered phase factor in (45) is found to be

${\displaystyle <0|T\{\exp [ie\int_{z}^{x}\,dx_{\mu}'A^{\mu (0)}(x')]\}|0>}$

$={\displaystyle \exp\{\frac{-ie^2}{2}[\int_{0}^{x}\,dx_{\mu}'\int_{0}^{x}
\,dx_{\nu}''D_{F}^{\mu \nu}(x'-x'') + \int_{z}^{0}\,dx_{\mu}'\int_{z}^{0}
\,dx_{\nu}''D_{F}^{\mu \nu}(x'-x'')]}$
\begin{equation}
-ie^2\int_{0}^{x}\,dx_{\mu}'\int_{z}^{0}\,dx_{\nu}''
D_{F}^{\mu \nu}(x'-x'')\} \,\,,
\end{equation}
where $x$ and $z$ are given by (46).

The first and second terms in the argument of the exponential in the last
equation can be cast into $G^{c}(p)$ and $G^{c}(p')$, respectively, while the
third one together with $\beta_{\mu}$ is not but the vertex function, which,
after performing the path integrations reads
\[{\cal B}_{\mu}(p,p')=\beta_{\mu}\exp \{\frac{-ie^2}{2} \int \, \frac{d^4k}
{{(2\pi)}^4}\,\frac{1}{(p.k)(p'.k)}p_{\mu}D_{F}^{\mu \nu}(k)p_{\nu}'\]
\begin{equation}
\times [(e^{-i\frac{x_0}{p_0}k.p}-1)(1-e^{i\frac{z_0}{p_0'}k.p'})+
(e^{i\frac{x_0}{p_0}k.p}-1)(1-e^{-i\frac{z_0}{p_0'}k.p'})]\}\,\,,
\end{equation}
where we have used (36) again. In the asymptotic limit
$z_0 \rightarrow -\infty$, $x_0 \rightarrow \infty$ the path-ordered phase
factor in (48) reduces to
\begin{equation}
\exp\{ie^2\int\,\frac{d^4k}{{(2\pi)}^4}\frac{p.p'}{(k^2+i\epsilon)
(k.p)(k.p')}\}
\end{equation}
in the Feynman gauge. This coincides with the infrared singular exponential
which factorizes in the vertex function for both spinorial and the usual
scalar QED. In the former it is obtained via perturbation theory${}^{[3]}$,
while in the later through an alternative laborious method${}^{[12]}$.

\vspace{1.0cm}

{\large\bf 5.\, Concluding Remarks}
\vspace{0.5cm}

We have obtained the meson propagator and the vertex function for the
linearized version of scalar QED in the infrared approximation, using the
Kubo's technique. In order to make it applicable, the effective interaction
Hamiltonian was written as a product of the mesonic current and the
electromagnetic field, thus eliminating the quartic interaction term in the
S-matrix of the original theory. This also enables us to investigate the
causal structure of the S-matrix${}^{[14]}$ in the effective theory, and the
infrared problem in the adiabatic limit${}^{[15]}$ as well.

We have shown that the infrared singularities can be extracted as
exponential phase factors from the two- and three-point Green's functions in
momentum space. The infrared phase factor (38) in the meson propagator
coincides with the divergent exponential factor in the fermion propagator in
the Kulish-Faddeev model${}^{[6]}$ for $y_0 \rightarrow -\infty$,
$x_0 \rightarrow \infty$. On the other hand, at asymptotic times, the
path-ordered phase factor (49) in the Fourier transform of
\[G_{\mu}(x,0,z)=<0|T\{\psi(x):\overline{\psi}(0)\beta_{\mu}\psi(0):\overline
{\psi}(z)|0>\]
also coincides with the resulting exponentiation of infrared divergences in
the Grammer-Yennie perturbation theory. Therefore, we are led to conclude
that spin effects do not manifest in the scattering amplitudes in the
infrared region${}^{[5],[13]}$, as expression (29) for the asymptotic current
of charged particles suggests.

In addition, equation (41) permits one to extend the analisis of infrared
divergences in dimensions other than three${}^{[11]}$. This requires closer
examination, to be presented in an opportune occasion. Finally, it is
noteworthy the remarkable simplicity of the method, which naturally
generalizes to the calculation of n-point Green's functions for interactions
of type (17), in contrast with the above mentioned methods.

\newpage

{\large\bf 6.\, References}
\vspace{0.5cm}

\begin{description}
\item[{[1]}] {\sf F. Bloch and A. Nordsieck, Phys. Rev.} {\bf 52} {\sf (1937)
54;}

\item[{[2]}] {\sf D. Yennie, S. Frautschi and H. Suura, Ann. Phys.} {\bf 13}
{\sf (1961) 379;}

\item[{[3]}] {\sf G. Grammer and D. R. Yennie, Phys. Rev.} {\bf D 8} {\sf
(1973) 4332;}

\item[{[4]}] {\sf V. Chung, Phys. Rev.} {\bf B 140} {\sf (1965) 1110;}

\item[{[5]}] {\sf P. Kulish and L. Faddeev, Theor. Math. Phys.} {\bf 4} {\sf
(1971) 745;}

\item[{[6]}] {\sf D. Zwanziger, Phys. Rev.} {\bf D 11} {\sf (1975) 3481, 3504;}

\item[{[7]}] {\sf R. Kubo, Prog. Theor. Phys.} {\bf 66} {\sf (1981) 1816;}

\item[{[8]}] {\sf T. Kinoshita, Prog. Theor. Phys.} {\bf 5} {\sf (1950) 473;}

\item[{[9]}] {\sf N. Kemmer, Proc. Roy. Soc.} {\bf A 173} {\sf (1939) 91;}

\item[{[10]}] {\sf A. V. Svidzinskii, J. Exptl. Theoret. Phys.} {\bf 31}
{\sf (1956) 324;}

\item[{[11]}] {\sf B. M. Pimentel and J. L. Tomazelli, J. Phys.} {\bf G 20}
{\sf (1994) 845;}

\item[{[12]}] {\sf M. C. Berg\`{e}re and L. Szymanowski, Phys. Rev.} {\bf D 26}
{\sf (1982) 3550;}

\item[{[13]}] {\sf D. Rello, Phys. Rev.} {\bf D 29} {\sf (1984) 2282;}

\item[{[14]}] {\sf G. Scharf, ``Finite Quantum Electrodynamics'', 2nd. ed.,
Springer-Verlag, Berlin, to appear;}

\item[{[15]}] {\sf M. D\"utsch, F. Krahe and G. Scharf, J. Phys.} {\bf G 19}
{\sf (1993) 485, 503;}
\end{description}

\end{document}